\documentstyle[aps,prl,twocolumn,epsf,floats]{revtex}

\begin{document}

\draft

\title{Interlayer Magnetic Frustration in Quasi-stoichiometric Li$_{1-x}$Ni$_{1+x}$O$_{2}$}

\author{E. Chappel, M. D. N\'{u}\~{n}ez-Regueiro, S. de Brion, and G. Chouteau}
\address{Grenoble High Magnetic Field Laboratory, MPI-FKF and CNRS, B.P. 166, 38042 Grenoble Cedex 9, France\\}
\author{V. Bianchi, D. Caurant and N. Baffier}
\address{Laboratoire de Chimie Appliquée de l'Etat Solide, ENSCP, 11 rue Pierre et Marie Curie, 75231 Paris Cedex 5, France\\}

\date{\today}

\maketitle

\begin{abstract}
Susceptibility, high-field magnetization and submillimeter wave electron spin resonance
measurements of layered quasi-stoichiometric  Li$_{1-x}$Ni$_{1+x}$O$_{2}$ are reported
and compared to isomorphic NaNiO$_{2}$. A new mechanism of magnetic frustration
induced by the excess Ni ions always present in the Li layers is proposed. We finally
comment on the possible realization of an orbital liquid state in this controversial
compound.
\end{abstract}

\pacs{75.40.Cx, 76.30.-v, 75.10.Jm}

Since its synthesis in 1958 by Goodenough et al. \cite{good}, LiNiO$_{2}$ is a subject
of continuous debate. Its ideal structure can be described as a packing of Li and
NiO$_{2}$ slabs built up of edge sharing NiO$_{6}$ octahedra. Therefore, magnetic Ni and
nonmagnetic Li hexagonal planes alternate along the <111> direction, giving rise to a
quasi-2D magnetic lattice. Detailed chemistry  analysis of this compound has been
motivated by its potential application in rechargeable batteries; this allowed to overcome
the initial disagreement between results from different groups due to the sensitivity
of the physical properties to the sample preparation method. On the other hand,
theoretical interest on this system comes from the interplay between different
degrees of freedom : doubly degeneracy of the Ni$^{3+}$ (t$_{2g}^{6}$e$_{g}^{1}$)
orbitals and their eventual coupling to the S=1/2 spins, the effect of frustration in the
triangular Ni lattice, the elusive nature of the magnetic interactions.
In spite of numerous studies and significant progress on these subjects \cite{tokura},
the puzzling absence of  both orbital and magnetic ordering, indeed clearly
observed in isomorphic NaNiO$_{2}$ \cite{chappel,chappelbis}, remains a mystery.
More recently, LiNiO$_{2}$ has been considered as the first realization of a quantum
spin orbital liquid \cite{feiner,li,kitaoka,mila}.

Here we report new measurements on well characterized homogeneous
Li$_{1-x}$Ni$_{1+x}$O$_{2}$. We study three samples, one of them being, to our
knowledge, the closest one to stoichiometry reported up to now, and we compare
their behavior to NaNiO$_{2}$. The overall results agree with recent theoretical
development \cite{mostovoy} concerning the decoupling of the orbital and
spin degrees of freedom, particular to these frustrated
Jahn-Teller (JT) systems, and the always ferromagnetic (FM) sign of the
intralayer Ni-Ni magnetic interactions. They also confirm the cluster model that
we have proposed \cite{nunez} to describe the dependence on the concentration
$x$ of different properties. Furthermore, we can conclude that most probably
stoichiometric LiNiO$_{2}$ does not exist, and that the clusters formed around
the excess Ni ions on Li planes are responsible for the peculiar behavior
of this system. In fact, when the intrinsic weak antiferromagnetic (AF)
interaction between adjacent Ni layers, neglected in previous theoretical
works \cite{mostovoy,nunez}, is taken into account, the effective FM coupling
induced by those clusters frustrates the AF stacking of the FM Ni planes,
hindering the long range 3D-magnetic ordering observed in NaNiO$_{2}$ below
20K \cite{chappelbis}.

The detailed description of the synthesis conditions and of the structural
characterization of our NaNiO$_{2}$ and Li$_{1-x}$Ni$_{1+x}$O$_{2}$ samples
has been given elsewhere \cite{chappelbis,bianchi}. The amount of extra Ni
ions were determined by Rietveld refinement : $x$=0.004, 0.016 and 0.060
($\pm$0.002) in the latter system.
The magnetic susceptibility measurements were performed on a SQUID magnetometer under
1mT field, between 2 and 300K using pressed pellets. High magnetic fields up to 23T
were obtained using resistive magnets. Electron spin resonance (ESR) measurements
were carried out at different frequencies and temperatures, using Gunn oscillators
and Carcinotrons. The magnetic field (up to 12T) was produced by a superconducting
magnet.

Fig.\ \ref{fig1} reports the temperature dependence of the susceptibility M/H at 1mT,
showing the high quality of our Li$_{1-x}$Ni$_{1+x}$O$_{2}$ sample, $x$=0.004.
The curve presents a cusp around T$_{\chi}$=7.5K when heating after zero field
cooling (ZFC), whereas on field cooling (FC) M/H remains almost constant below T$_{\chi}$.
This is a typical spin-glass like behaviour. The inverse susceptibility H/M, shown in
the inset of Fig. 1, can be fitted with a Curie-Weiss law between 70 and 300K, with
an effective moment $\mu_{eff}$=2$\mu_{B}$ and a Weiss temperature $\theta$=+26K.
These results are similar to those of Yamaura et al.\cite{yamaura} with
T$_{\chi}$=8.5K, $\mu_{eff}$=1.91$\mu_{B}$ and $\theta$=+29.5K. The positive
sign of $\theta$ indicates the predominance of FM interactions and the effective
moment corresponds unambiguously to the low spin state (t$_{2g}^{6}$e$_{g}^{1}$)
of Ni$^{3+}$ with S=1/2. It is interesting to point out that, the most
stoichiometric sample has the lowest Weiss temperature, the linear dependence
of $\theta(x)$ for $x\rightarrow0$, precising the concentration determined from Rietveld
analysis \cite{bianchi}. This allows us to say that our $x\simeq$0.004 sample
is the most diluted reported to date. The existence of a well defined
susceptibility peak is an indication of the homogeneity of the sample.
Besides, if the Ni$^{2+}$ ions were not distributed at random in the Li layers,
this would lead to ferrimagnetic domains and to an additional anomaly at $\sim$240K
in the susceptibility (only for mT fields).

    Fig.\ \ref{fig2}  shows the magnetization at 4K of NaNiO$_{2}$ and Li$_{1-x}$Ni$_{1+x}$O$_{2}$
samples. The low field curvature, signature of ferrimagnetic clusters, vanishes when
$x\rightarrow$0, approaching the linear behavior observed \cite{chappelbis} for the
$A$-type antiferromagnet NaNiO$_{2}$. For this compound, the saturation is reached
at 10T and the moment 1$\mu_{B}$/Ni agrees again with the low spin state of Ni$^{3+}$.
For the Li$_{1-x}$Ni$_{1+x}$O$_{2}$ samples the saturation is almost achieved at 23T
with M$_{Sat}$=(1-$x$)$\mu$(Ni$^{3+})$. The Ni$^{2+}$ moments are not involved in the
latter formula since, they are still AF coupled, even at 23T, following the
Goodenough-Kanamori-Anderson (GKA) rules (Ni$^{2+}$-O-Ni$^{3+}$ form 180$^{\circ}$
bonds) \cite{nunez}. The high field dependence of the magnetization of
Li$_{1-x}$Ni$_{1+x}$O$_{2}$ samples confirms the presence of extra AF
interactions, compared to NaNiO$_{2}$. Thus, the evolution of the
magnetization curve with concentration $x$ comes from ferrimagnetic clusters,
created by these extra Ni$^{2+}$ ions inducing AF couplings. Also the Arrot's
plots at various temperatures in Ref. \cite{bianchi} are typical of an AF
like NaNiO$_{2}$. In fact, the small number of clusters for this $x$ cannot explain
by itself such macroscopic AF behaviour, strongly suggesting that, also in
this case, there is an intrinsec weak AF interaction between adjacent Ni layers,
$\sim$1K in NaNiO$_{2}$ \cite{chappelbis}.

Two different ESR studies have been already performed on diluted
Li$_{1-x}$Ni$_{1+x}$O$_{2}$. Ohta et al. \cite{ohta}, by measurements between 1.8
and 256K, up to 370GHz with pulsed magnetic fields, have shown the existence of two
lines, already present at 86K, which split progressively with further cooling.
At 1.8K, the frequency vs. magnetic field diagram suggested the existence of an internal
field, since the extrapolated lines did not cross the origin. Barra et al. \cite{barra}
have performed similar ESR measurements at 73 and 246GHz, interpreting the splitting of
these two lines, observed below 130K, in terms of anisotropic $g$ factors with
$g_{\perp}$<$g_{\parallel}$. Then, they proposed a dynamical JT effect of the
Ni$^{3+}$ ions, which becomes static below 130K. Since $g_{\perp}$<$g_{\parallel}$,
the occupied orbital should be $\mid d_{x^2-y^2}>$, in contradiction with the
EXAFS study by Rougier et al. \cite{rougier}. In fact, these authors have
shown the existence of elongated octahedra from 300K, i.e. a local JT effect,
favoring the occupation of the $\mid d_{3z^2-r^2}>$ orbital, as for isomorphic
NaNiO$_{2}$. In order to clarify this point we have undertaken a complete ESR study of
the most stoichiometric sample, up to 285GHz at various temperatures.
A typical low-temperature spectra in shown in the inset of Fig.\ \ref{fig3}.
The signal is strongly anisotropic with the splitting discussed above,
showing striking analogies with NaNiO$_{2}$ (Figs.8 and 9 in Ref.\cite{chappel}).
The frequency evolution at 50K of these two features, called A and B, are presented
in Fig.\ \ref{fig3}. The extrapolated lines cross the frequency axis at 3$\pm$1GHz, -1$\pm$1,
for A and B respectively, confirming the existence of an internal field and invalidating
the static-dynamical JT analysis. Such similar phenomenon in both,
the Na and Ni compounds, indicates that the splitting of the ESR lines
at low-temperature is not due to the JT effect but has a magnetic
origin. We correlate this effect with the deviation from the Curie-Weiss law,
observed below 70K for our most diluted sample (inset Fig.\ \ref{fig1}).

Fig.\ \ref{fig4} shows the qualitative different ESR spectra of quasi-stoichiometric
Li$_{1-x}$Ni$_{1+x}$O$_{2}$ compared to NaNiO$_{2}$ at T=200K, a temperature half lower
than the temperature at which the JT distortion takes place for NaNiO$_{2}$
(T$_{JT}\simeq$480K). Even at the high frequencies (230-285GHz) necessary to clearly
separate the two $g$ components in  NaNiO$_{2}$, there is now a single isotropic line
around $g$=2.17. We will come back to this result.

In  Li$_{1-x}$Ni$_{1+x}$O$_{2}$ there are only 90$^{\circ}$ Ni-O-Ni paths
within the same plane, while Ni$^{2+}$ ions in the Li planes introduce 180$^{\circ}$
Ni-O-Ni bonds. Considering the GKA rules and fitting the high temperature
susceptibility as a function of the concentration $x$ by a mean-field approach,
we have shown \cite{nunez} that the intralayer interactions are FM and the Ni-O-Ni
couplings induced by the extra Ni in Li layers are AF. Moreover,
in order to obtain a good description of the whole data, it was necessary
to take into account that the intralayer Ni$^{3+}$-O-Ni$^{2+}$
(Ni$^{2+}$ ions in the Ni layers due to charge compensation)
coupling $J_{F1}$ is stronger than the Ni$^{3+}$-O-Ni$^{3+}$ $J_{F2}$ one.
Though, ferrimagnetic cluster are formed : each Ni$^{2+}$ ion in the
Li plane connects 6 FM Ni ions (since the AF 180$^{\circ}$ Ni-O-Ni
coupling applies twice), i.e. 3 in each of the two adjacent Ni layers.
If also we consider the FM $J_{F1}$ between  Ni$^{2+}$ and Ni$^{3+}$
in the same Ni layer, then for $x$=1/12 homogeneously distributed excess
Ni$^{2+}$ ions, all sites are coupled. This percolation threshold
appears to be in agreement with experiments (see Fig. 2 in Ref.\cite{nunez}).

Since both intralayer interactions are FM, frustrated AF models in the
triangular Ni layer, yielding a spin liquid \cite{hirakawa,hirota,mertz} are
inappropriate for this system. Other works \cite{feiner,li,kitaoka,mila,reynaud}
proposed LiNiO$_{2}$ as an example of a quantum orbital spin liquid, to explain both
the absence of magnetic and orbital ordering. All these models assume a strong coupling
between the orbital and spin degrees of freedom [so-called Kugel-Khomskii (KK) systems
\cite{kugel}]. Surprisingly, no author comments the EXAFS results by
Rougier et al. \cite{rougier} showing the existence of two different Ni-O distances
in quasi-stoichiometric Li$_{1-x}$Ni$_{1+x}$O$_{2}$, with 4 short and 2 long
Ni-O bonds, as in NaNiO$_{2}$. Even if there is no macroscopic structural transition,
this elongation of the NiO$_{6}$ octahedra occurs even at room temperature, and the
local JT effect of the Ni$^{3+}$ ions appears as the relevant process for the breakdown
of orbital degeneracy. The KK rules concern cubic lattices with 180$^{\circ}$ bonds, and
conclusions for these frustrated JT systems are different \cite{mostovoy}. In fact,
as we have first point out \cite{nunez}, at 90$^{\circ}$, and independently of the
orbital e$_{g}$ occupation, the intralayer magnetic Ni-Ni coupling is always FM.

According to the experimental results and theoretical considerations
discussed above, we propose here a new mechanism of magnetic frustration in
diluted Li$_{1-x}$Ni$_{1+x}$O$_{2}$. The pure compound LiNiO$_{2}$,
if it exists would have the magnetic structure of NaNiO$_{2}$. The magnetization,
susceptibility and ESR measurements clearly show that the magnetic properties of
Li$_{1-x}$Ni$_{1+x}$O$_{2}$ approach those of NaNiO$_{2}$ when the excess Ni
concentration $x\rightarrow0$. At low temperature, the weak AF interplane
Ni-O-Li-O-Ni interaction of the NaNiO$_{2}$ type \cite{chappelbis} should lead to
an AF macrocopic order (AF alternation of adjacent FM Ni planes), in agreement with
Arrott's plots and magnetization curves. However, it seems
that the smaller size of the Li ions does not allow the perfect stacking of the
Na-Ni layers, and that a Ni concentration $x\neq$0 always goes into the
Li layers, inducing effective interplane Ni-Ni local FM couplings. Such
competition of interactions leads to magnetic frustration, prevents the stabilization
of long range ordering, and explains the spin glass behaviour observed in
quasi-stoichiometric Li$_{1-x}$Ni$_{1+x}$O$_{2}$ at low temperature, without the
necessity of evoking a spin liquid state. Fig.\ \ref{fig5} shows a sketch of this frustration
mechanism. Assuming that spins around a cluster turn progressively like
a magnetic wall, to finally adopt the AF stacking of NaNiO$_{2}$, an estimation
of the number of perturbed spins by each additional Ni ion can be made. To simplify,
in the case of an uniaxial crystal, the characteristic wall length writes
\begin{equation}
\delta_{0}=a\sqrt{\frac{8H_{E}}{3H_{A}}}
\end{equation}
where $a$ is the cell parameter, $H_{E}$ and $H_{A}$ are the exchange and
anisotropic fields, respectively \cite{herpin}. Taking the characteristic field values
obtained  \cite{chappelbis} for NaNiO$_{2}$, yields $\delta_{0}=6a$. In our hexagonal
symmetry, up to sixth-neighbor spins are perturbed, i.e. $\sim$60 spins
per cluster. Therefore, less than 1$\%$ of excess Ni in the Li planes can induce
complete magnetic disorder in Li$_{1-x}$Ni$_{1+x}$O$_{2}$.

EXAFS data indicate a JT elongation of the NiO$_{6}$ octahedra \cite{rougier},
but this effect remains local : there is no macroscopic structural transition.
It is noteworthy that the isotropic ESR line at 200K is not necessary in contradiction
with EXAFS. For instance, a configuration with elongated octahedra in orthogonal
directions $\mid d_{3z^2-r^2}>$, $\mid d_{3x^2-r^2}>$, $\mid d_{3y^2-r^2}>$ could
explain both data. On the other hand, recent theoretical work \cite{mostovoy}
proposed that the high degeneracy of mean-field orbital ground state in the
frustrated Ni lattice is lifted by quantum orbital fluctuations (Villain's order by
disorder), which select particular ferro-orbital states, as observed in NaNiO$_{2}$.
Then a weaker electron-lattice coupling, not enough for the stabilization of
that ferro-orbital state, or even stronger orbital fluctuations going against it, were
invoked to explain the absence of a cooperative JT effect in LiNiO$_{2}$.

Finally, we would like to point out that the susceptibility anomaly at
T$_{of}\simeq$400K reported by Reynaud et al.\cite{reynaud}, below which they claim
that an orbitally frustrated state is established, is very weak and within the
experimental error ($\Delta\chi\sim$3x10$^{-5}cm^{3}$/mol).
Therefore, up to now, only the absence of a macroscopic structural distortion and
the isotropic shape of the ESR line (Fig.\ \ref{fig4}) are indications of this complicate
orbital state. Orbitals are much more difficult to measure than spins, and it will
be a delicate task to distinguish between ordered chains in orthogonal directions,
disordered or fluctuating orbitals. In any case this will not influence the magnetic
behavior, that can be independently explained by the new frustration mechanism of Fig.\ \ref{fig5}.
Small angle neutron measurements are in progress, in order to "see" these
ferrimagnetic clusters.

In conclusion, from low and high magnetic field and ESR measurements on
homogeneous quasi-stoichiometric Li$_{1-x}$Ni$_{1+x}$O$_{2}$, we have shown
that the extra Ni ions always present in the Li planes, induce magnetic frustration
in the low $x$ limit, and this can explain the unusual magnetic properties.
We have shown that the splitting of the ESR lines below the temperature at which
the susceptibility deviates from the Curie-Weiss law, has a magnetic origin.
The single isotropic ESR line above this temperature, together with EXAFS data,
indicate a peculiar orbital occupation with elongated octahedra.

We are indebted to A. Sulpice for the low temperature magnetization measurements.

\begin{figure}[tb]
\caption{Temperature dependence of M/H at 1mT for  Li$_{1-x}$Ni$_{1+x}$O$_{2}$,
$x$=0.004. The open and full circles correspond to measurements on FC and on ZFC, respectively.
Inset: H/M vs. T at 1mT for the same sample; the continuous line shows the Curie-Weiss law.}
\label{fig1}
\end{figure}

\begin{figure}[tb]
\caption{Magnetization up to 23T of NaNiO$_{2}$ and various
quasi-stoichiometric Li$_{1-x}$Ni$_{1+x}$O$_{2}$ samples at 4K.}
\label{fig2}
\end{figure}

\begin{figure}[tb]
\caption{From typical low temperature ESR spectra of Li$_{1-x}$Ni$_{1+x}$O$_{2}$,
$x$=0.004, showing two  features A and B (see example in the inset),
very similar to NaNiO$_{2}$, the frequency-field diagram is proposed.}
\label{fig3}
\end{figure}

\begin{figure}[tb]
\caption{High temperature (T=200K) frequency-field ESR diagrams for NaNiO$_{2}$ (dashed line)
and Li$_{1-x}$Ni$_{1+x}$O$_{2}$, $x$=0.004 (continuous line).
Insets : typical spectra for both samples.}
\label{fig4}
\end{figure}

\begin{figure}[tb]
\caption{Sketch of the magnetic frustration mechanism proposed here for quasi-stoichiometric
Li$_{1-x}$Ni$_{1+x}$O$_{2}$. The effective FM interplane coupling
induced by extra Ni ions in the Li planes, hiddens the macroscopic magnetic
ordering of NaNiO$_{2}$, driven by the weak AF interplane interaction.}
\label{fig5}
\end{figure}


\begin{references}

\bibitem[*]{byline} E-mail: nunezreg@polycnrs-gre.fr (MDNR)

\bibitem{good} J. B. Goodenough, D. G. Wickham, and W. J. Croft, J.\ Phys.\ Chem.\ Solids {\bf5}, 107 (1958).

\bibitem{tokura} Y. Tokura and N. Nagaosa, Science {\bf5465}, 462 (2000).

\bibitem{chappel} E. Chappel et al., Eur.\ Phys.\ J.\ B {\bf17}, 615 (2000).

\bibitem{chappelbis} E. Chappel et al., Eur.\ Phys.\ J.\ B {\bf17}, 607 (2000).

\bibitem{feiner} L. F. Feiner, A. M. Oles, and J. Zaanen, Phys.\ Rev. Lett. {\bf78}, 2799 (1997); and Phys.\ Rev.\ B {\bf61}, 6257 (2000).

\bibitem{li} Y. Q. Li, M. Ma, D. N. Shi, and F.-C. Zhang, Phys.\ Rev. Lett. {\bf81}, 3527 (1998).

\bibitem{kitaoka} Y. Kitaoka et al., J. \ Phys. Soc. Japan {\bf67}, 3703 (1998).

\bibitem{mila} M. van der Bossche, F.-C. Zhang, and F. Mila, Eur.\ Phys.\ J.\ B {\bf17}, 367 (2000).

\bibitem{mostovoy} M. V. Mostovoy and D. I. Khomskii, cond-mat/0201420 (2002).

\bibitem{nunez} M. D. N\'{u}\~{n}ez-Regueiro, E. Chappel, G. Chouteau, and C. Delmas,
Eur.\ Phys.\ J.\ B {\bf16}, 37 (2000).

\bibitem{bianchi} V. Bianchi et al., Solid State Ionics {\bf140}, 1 (2001).

\bibitem{yamaura} K. Yamaura et al., J. Solid State Chem. {\bf127}, 109 (1996).

\bibitem{ohta} H. Ohta et al., Physica B {\bf211}, 217 (1995), and {\bf237-238}, 64 (1997).

\bibitem{barra} A.-L. Barra et al., Eur.\ Phys.\ J.\ B {\bf7}, 551 (1999).

\bibitem{rougier} A. Rougier, C. Delmas, and A. V. Chadwick, Solid State Comm. {\bf94}, 123 (1995).

\bibitem{hirakawa} K. Hirakawa et al., J. Magn. Magn Mat. {\bf90-91}, 279 (1990), and
J. Phys. : Condens. Matter {\bf3}, 4721 (1991).

\bibitem{hirota} K. Hirota et al., J.\ Phys. Soc. Japan {\bf67}, 3703 (1998).

\bibitem{mertz} D. Mertz et al., Phys. Rev. B {\bf61}, 6257 (2000).

\bibitem{reynaud} F. Reynaud et al., Phys. Rev. Lett. {\bf86}, 3638 (2001).

\bibitem{kugel} K. I. Kugel and D. I. Khomskii, Sov.\ Phys.\ JETP {\bf52}, 501
(1981); Sov.\ Phys.\ Uspekhi {\bf25}, 232 (1982)

\bibitem{herpin} A. Herpin, Théorie du Magn$\acute{e}$tisme (Presse
Universitaire de France, 1968).

\end{references}
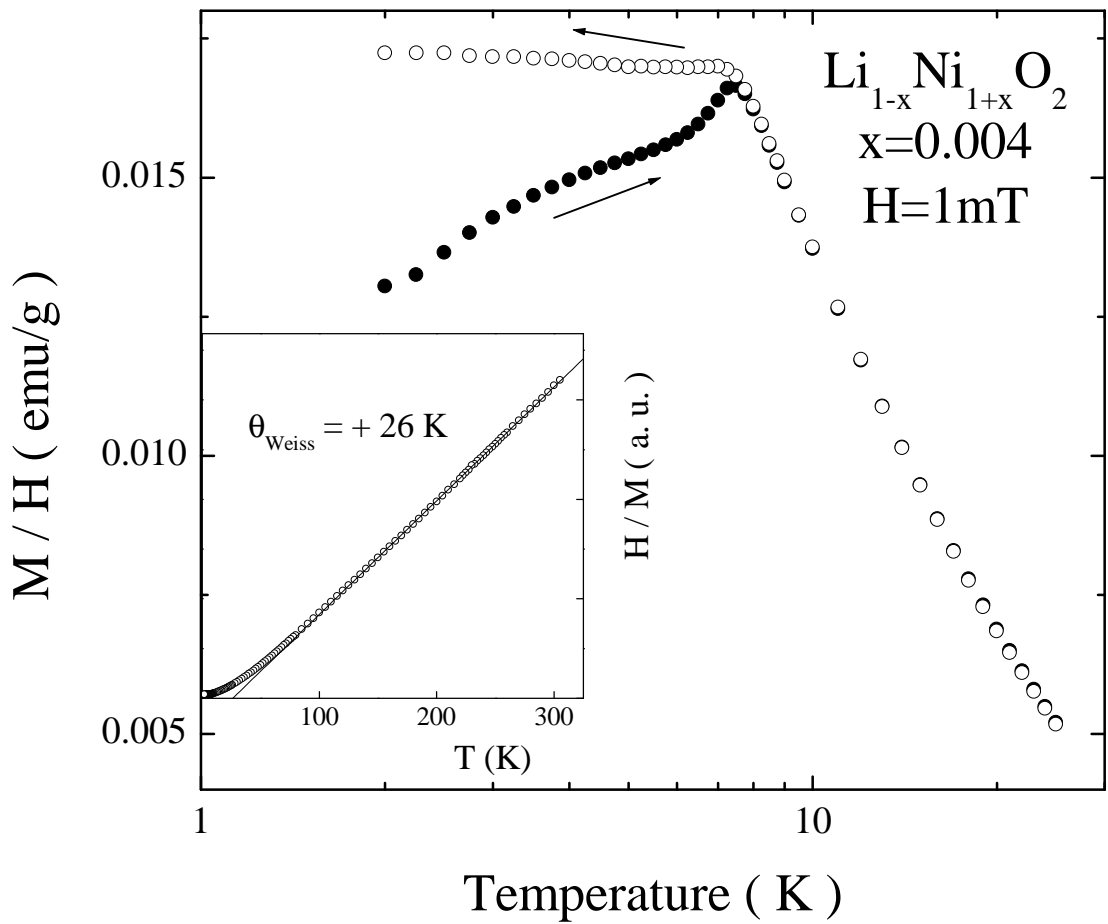
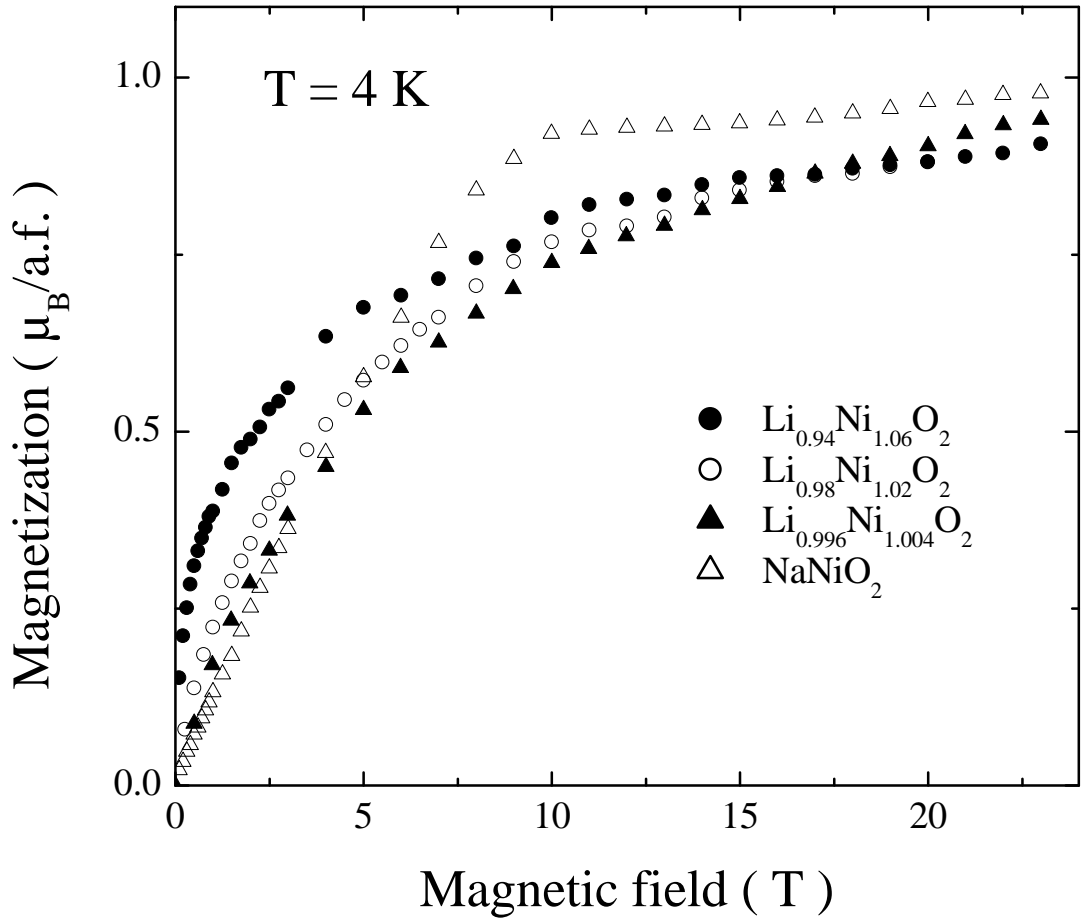
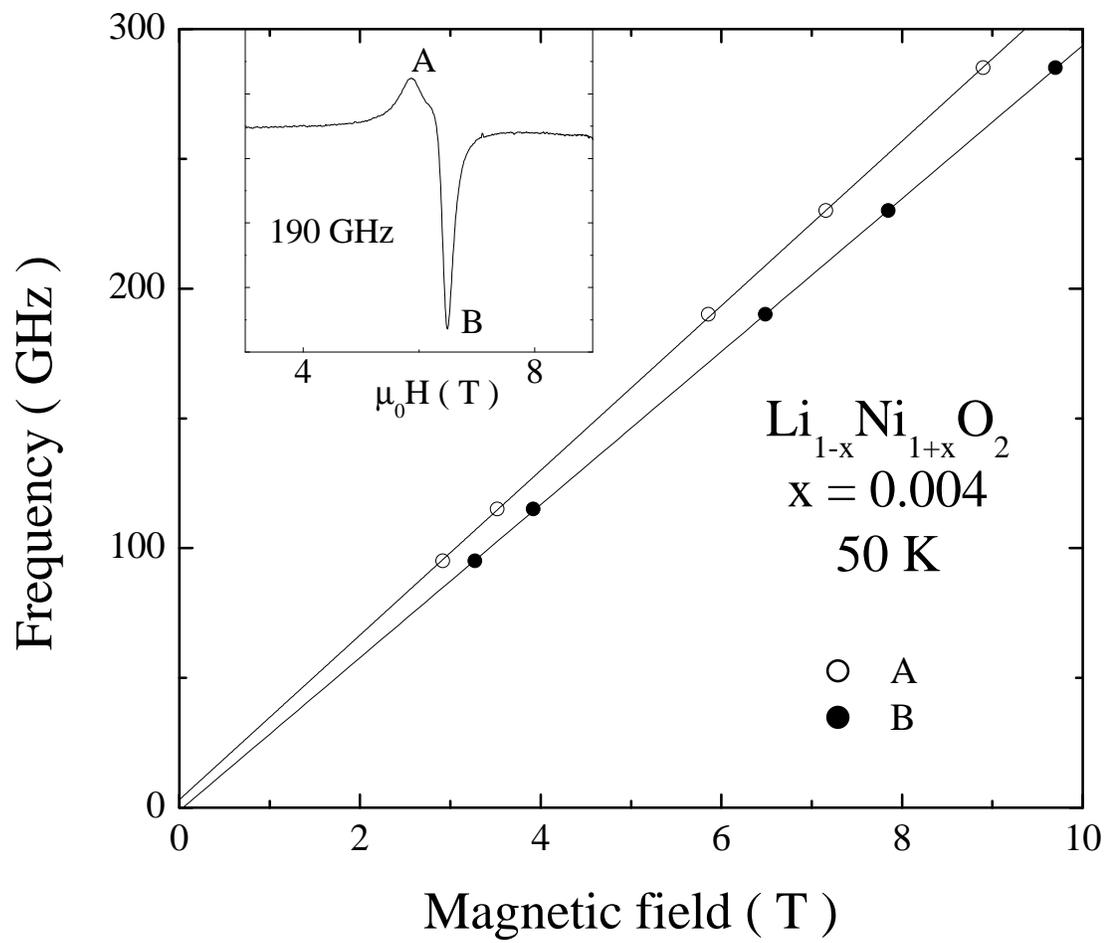
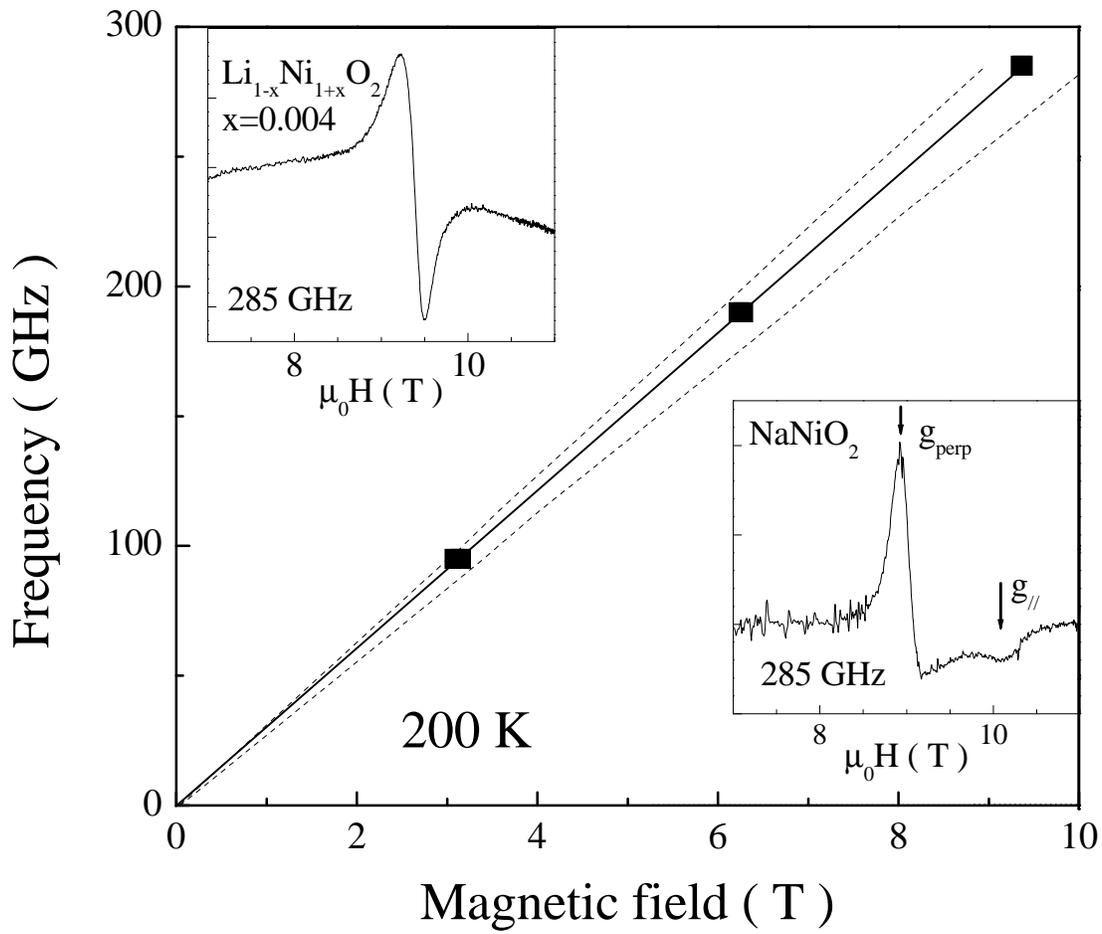
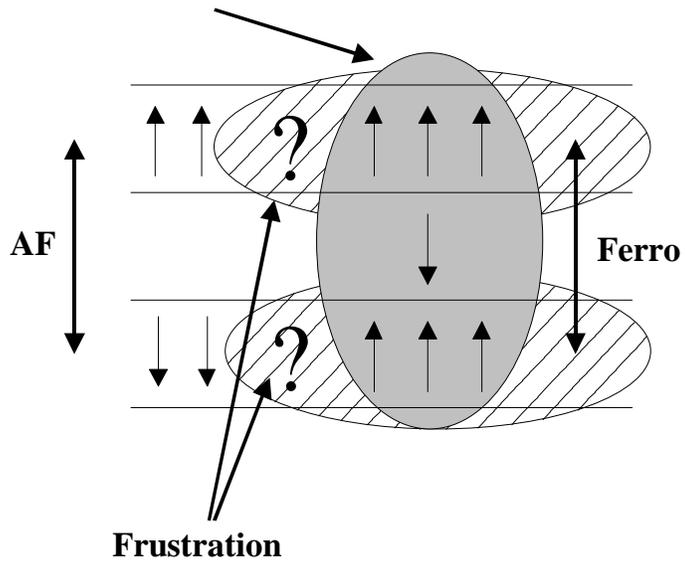
\end{document}